\documentclass[conference]{IEEEtran}
\IEEEoverridecommandlockouts

\usepackage{cite}
\usepackage{amsmath,amssymb,amsfonts}
\usepackage{xcolor}
\usepackage{graphicx}
\usepackage{textcomp}
\def\BibTeX{{\rm B\kern-.05em{\sc i\kern-.025em b}\kern-.08em
    T\kern-.1667em\lower.7ex\hbox{E}\kern-.125emX}}
\usepackage{comment}
\usepackage{dirtytalk}
\usepackage{amsmath}
\DeclareMathOperator{\Tr}{tr}
\begin{document}

\title{Untrained DNN for Channel Estimation of  RIS-Assisted Multi-User OFDM System with Hardware Impairments
\thanks{This work was supported by the Academy of Finland 6Genesis Flagship (grant no. 318927).}}

\author{\IEEEauthorblockN{Nipuni  Ginige, K. B. Shashika Manosha, Nandana Rajatheva, and Matti Latva-aho}\IEEEauthorblockA{Center for Wireless Communications,
University of Oulu,
Finland \\
\{nipuni.ginige,  nandana.rajatheva, matti.latva-aho\}@oulu.fi, manoshadt@gmail.com}}

\maketitle

\begin{abstract}  
Reconfigurable intelligent surface (RIS) is an emerging technology for improving performance in fifth-generation (5G) and beyond networks. Practically channel estimation of RIS-assisted systems is challenging due to the passive nature of the RIS. The purpose of this paper is to introduce a deep learning-based, low complexity channel estimator for the RIS-assisted multi-user single-input-multiple-output (SIMO)  orthogonal frequency division multiplexing (OFDM) system with hardware impairments. We propose an untrained deep neural network (DNN) based on the deep image prior (DIP) network to denoise the effective channel of the system obtained from the conventional pilot-based least-square (LS) estimation and acquire a more accurate estimation. We have shown that our proposed method has high performance in terms of accuracy and low complexity compared to conventional methods. Further, we have shown that the proposed estimator is robust to interference caused by the hardware impairments at the transceiver and RIS. 
\end{abstract}


\section{Introduction}
Reconfigurable intelligent surfaces (RIS) is a planar surface that consists of a larger number of passive reflecting elements which can  change  the phases of the incoming signals to improve the performance of the system. Channel estimation has a crucial role in terms of the performance of wireless communication systems. Channel estimation in RIS-assisted systems plays a more important role since the estimated channel coefficients are used to find the optimal reflection coefficients which can maximize the performance of the system. Moreover, channel estimation in a RIS-assisted system is more challenging compared to the conventional communication system due to the passive nature and the large number of elements in RIS.

Channel estimation of the RIS-assisted systems can be categorized mainly into two parts based on the configuration of the RIS such as semi-passive RIS (some active elements in RIS) and fully-passive RIS. Most of the existing literature of channel estimation for RIS-assisted system is based on fully-passive RIS since it is more energy-efficient and cost-effective compared to the semi-passive RIS \cite{9326394}. Authors in \cite{9039554, 8683663, 9090876}, proposed to estimate the channel sequentially by only turning on the corresponding RIS element in one time symbol. However, this ON/OFF method is sub-optimal and the authors in \cite{9053695}, proposed an optimal channel estimation method using a RIS reflection pattern which follows the rows of discrete Fourier transform (DFT) and orthogonal frequency division multiplexing (OFDM) symbols equal to the number of elements in RIS for training. The above method has been used by authors in \cite{8937491}, to estimate the channel of the RIS-assisted OFDM system. The research work mentioned above  considered the scenario which had ideal hardware in transceiver and RIS. Therefore, their results are asymptotic. However, the authors in \cite{ papazafeiropoulos2021intelligent} have studied channel estimation of the RIS-assisted system when there are impairments in the hardware of the transceiver and RIS.

Recently machine learning (ML) is playing a big role in wireless communication applications. Deep learning algorithms can be used to improve the accuracy of channel estimation in wireless communication systems. The authors in \cite{8949757}, proposed a low complexity channel estimation technique using deep learning algorithms for a massive multiple-input-multiple-output (MIMO) OFDM system which is robust to pilot contamination. Their deep neural network (DNN) is based on the deep image prior (DIP) network and they have proven that they can achieve minimum mean square error (MMSE) performance by using least square (LS) estimations with low complexity than the MMSE. They have denoised the received signal using untrained DNN based on DIP and used that denoised received signal for conventional LS estimation. Also, they have shown that denoising the received signal does not increase complexity due to the untrained nature of the DNN.  DIP was proposed by the authors in \cite{lempitsky2018deep}, to address image reconstruction problems such as denoising without the need to train a large data-set beforehand. This model was further optimized by the authors in \cite{heckel2018deep}.

Traditional pilot-based channel estimation is not suitable to make more reliable channel estimation for the RIS-assisted systems since the high training overhead and computational complexity due to the large number of passive elements in the RIS. The authors in \cite{9366894}, proposed data-driven non-linear solutions based on deep learning to approximate the globally optimal minimum mean square error (MMSE) channel estimator for the scenario which has ideal hardware in transceiver and RIS. However, this method requires a large number of parameters which need to train with large data sets.  

The main contribution of this paper is to introduce a deep learning-based, low complexity channel estimation technique for RIS-assisted multi-user single-input-multiple-output (SIMO)  OFDM system. The proposed channel estimator is robust to interference caused by the hardware impairments at the transceiver and RIS. In this paper, we propose a DNN which does not require any training, based on the DIP network to improve the estimation accuracy of the RIS-assisted multi-user SIMO  OFDM system. In this untrained DNN model, we optimize parameters of the neural network for each channel realization on the fly without training beforehand. We use a  RIS reflection pattern which follows the rows of DFT during the channel estimation phase. Our main idea is to denoise the conventional pilot-based LS estimation of the effective channel of the RIS-assisted system via our proposed DNN. Then use that denoised signal to find the user equipment (UE)- base station (BS) direct channel and the UE-RIS-BS cascade channels with the help of the knowledge of the reflection pattern. As the proposed DNN does not need any training, it does not increase the complexity or latency. However it  only causes significant improvement in the accuracy of the estimation even when there are impairments in hardware.

The rest of the paper is organized as follows. The system model and conventional channel estimation methods are described in Section \ref{SM}. In section \ref{DNN}, we present the proposed deep learning based channel estimator. The numerical results are presented in Section \ref{results} and Section \ref{conclusion} concludes our paper.

\section{System Model and Motivation}
\label{SM}
\subsection{System Model}
\begin{figure}[ht]  
	\centering
	\includegraphics[width=0.48\textwidth]{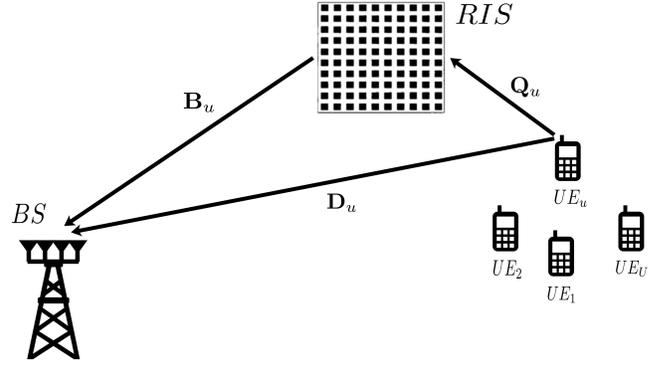}
	\caption{Illustration of the system model.}
	\label{fig:Illustration of the system model}
\end{figure}

We consider an uplink of a RIS-assisted multi-user SIMO   OFDM system as illustrated in Fig. \ref{fig:Illustration of the system model}. We assume that transmission is from $U$ single antenna UEs to a BS equipped with $K$ receive antennas.  RIS is composed of $M \times M =\Tilde{M} $ reflecting elements. We divide the RIS into $M$ sub-surfaces by grouping adjacent elements which have high correlation and assume that there is a common reflection coefficient within the $m$th sub-surface, where $m=1,\cdots,M$. There are $N$ sub-carriers in each OFDM symbol. At the UE side, OFDM symbol of each UE is denoted by ${\mathbf{x}_u}\in \mathbb {C}^{N \times 1} $.

Let ${\mathbf{D}}_{u} = \left[ \mathbf{d}_u[1],\mathbf{d}_u[2],\ldots,\mathbf{d}_u[N] \right]^{T}\in \mathbb{C}^{N \times K }$  denotes the frequency response of the direct UE-BS channel from $u$th UE.  Let ${\mathbf{Q}}_{u} = \left[ \mathbf{q}_u[1],\mathbf{q}_u[2],\ldots,\mathbf{q}_u[N] \right]^{T}  \in \mathbb{C}^{N \times M}$ and ${\mathbf{B}}_{u} =\left[ \tilde{\mathbf{B}}_u[1],\tilde{\mathbf{B}}_u[2],\ldots,\tilde{\mathbf{B}}_u[N]\right]^{T}  \in \mathbb{C}^{N \times K \times M}$  denote the frequency response of the UE-RIS channel and the frequency response of the RIS-BS channel associated to $u$th UE, respectively. Let $\tilde{\mathbf{G}}_u[n]=\tilde{\mathbf{B}}_u[n] \text{diag}(\mathbf{q}_u[n])\in \mathbb{C}^{K\times M}$ denotes the frequency response of the cascaded channel between UE-RIS-BS associated with $u$th UE, in $n$th sub-carrier, without the effect of phase shift at the RIS.  The BS and RIS lie at fixed locations and we consider low-mobility UEs. Therefore we assume quasi-static block fading for all the links which means the channel is constant in one transmission frame.

The frequency domain  received signal at the BS is given by

\begin{equation} \label{eq1}{\mathbf{Y}}=\sum _{u=1}^{U} {\mathbf{X}_u} \left( { \mathbf{G}_u \mathbf{\phi}  + {\mathbf{D}}_{u}} \right) + {\mathbf{V}},\end{equation}
where  ${\mathbf{X}_u}=\text {diag} \left ({{\mathbf{x_u}}}\right)$ is the diagonal matrix of the transmitted OFDM symbol ,  ${\mathbf{G}}_{u} = \left[ \tilde{\mathbf{G}}_u[1],\tilde{\mathbf{G}}_u[2],\ldots,\tilde{\mathbf{G}}_u[N] \right]^{T}\in \mathbb{C}^{N \times K \times M}$ is the cascade channel between UE-RIS-BS related to $u$th UE, ${\mathbf{V}} \sim {\mathcal N_{c} }({\mathbf{0}} , \sigma ^{2}{\mathbf{1}}_{N \times K})$ denotes the additive white Gaussian noise (AWGN) matrix with noise power of $\sigma^2$. Further $\mathbf{\phi} = [\phi _{1},\phi _{2}, \ldots \phi _{M}]^T \in \mathbb{C}^{M \times 1}$ is the reflection coefficient vector of the RIS  and $\phi _{m}= \beta_m e^{-j\varphi _{m}}$ denotes the reflection coefficient of $m$th sub-surface where $\beta _{m}\in (0, 1]$ is the reflection amplitude and $\varphi _{m}\in (0, 2\pi]$ is the phase shift. During the channel estimation process we fix $\beta _{m}=1$,  $\forall \; m$ and only adjust the phase shift $\varphi _{m}$.  Therefore, effective channel of the RIS-assisted system related to $u$th UE is ${ \mathbf{H}_u = \mathbf{G}_u \mathbf{\phi}  + {\mathbf{D}}_{u}}$. We do not estimate $\mathbf{Q}_{u}$ and ${\mathbf{B}}_{u}$ separately  since estimating the cascaded channel ${\mathbf{G}}_{u}$, is sufficient for applications\cite{9053695}. 

\subsection{Channel Model}

The time domain channels of all links can be described in terms of correlated Rayleigh fading distributions as follows:
\begin{align}
\mathbf{a}_{u, l}^d = \sqrt{\beta_{u, l}^d}\mathbf{C}_{BS, u}^{1/2}\mathbf{z}_{u, l}^d, \\
\mathbf{a}_{u, l}^q = \sqrt{\beta_{u, l}^q}\mathbf{C}_{RIS, u}^{1/2}\mathbf{z}_{u, l}^q, \\
\mathbf{A}_{u, l}^b = \sqrt{\beta_{u, l}^b}\mathbf{C}_{BS, u}^{1/2}\mathbf{Z}_{u, l}^b\mathbf{C}_{RIS, u}^{1/2},
\end{align}
where $\mathbf{a}_{u, l}^d,\mathbf{a}_{u, l}^q$ and $\mathbf{A}_{u, l}^b$ is the corresponding time domain channel of UE-BS, UE-RIS and RIS-BS links respectively. Also, $\beta_{u, l}^d, \beta_{u, l}^q$ and $\beta_{u, l}^b$ are the path-loss corresponding to the direct UE-BS, UE-RIS and RIS-BS paths. Furthermore, $\mathbf{C}_{BS ,u} \in \mathbb{C}^{K \times K}$   and $\mathbf{C}_{RIS, u} \in \mathbb{C}^{M \times M}$ denotes the spatial correlation matrices at the BS and RIS respectively with $\Tr({\mathbf{C}_{BS, u}})=K$ and $\Tr({\mathbf{R}_{RIS, u}})=M$. Spatial correlation model for RIS generates using \cite{bjornson2020rayleigh}. Moreover, $\mathbf{z}_{u, l}^d  \sim {\mathcal N_{c} }({\mathbf{0}} , {\mathbf{I}}_{K}) ,\mathbf{z}_{u, l}^q   \sim {\mathcal N_{c} }({\mathbf{0}},{\mathbf{I}}_{M}) $ and $ \mathbf{Z}_{u, l}^b \sim {\mathcal N_{c} }({\mathbf{0}} , {\mathbf{1}}_{K\times M})$ denotes the corresponding Raleigh fading matrices.

The channel response in the frequency domain at $n$th sub-carrier for the direct UE-BS link can be  written as
\begin{equation} \label{ch} {\mathbf {d}}_u[n]=\sum _{l=0}^{ L_d} \mathbf {a}_{u, l}^d  e^{- {\mathrm {j}}\frac {2\pi n}{N}l} ,
\end{equation}
where $L_d$ is the number of delay taps  in the direct UE-BS link.  We can write the channel response in the frequency domain at $n$th sub-carrier for the UE-RIS link and RIS-BS similar as in \eqref{ch} by replacing ${a}_{u, l}^d$  with ${a}_{u, l}^q$, ${A}_{u, l}^b$ and $L_d$ with $L_q$, $L_b$, respectively. Here $L_q$ and $L_b$ are the number of delay taps in UE-RIS link and RIS-BS link, respectively.

 \subsection{Channel Estimation-LS}
 \label{LS}
We consider one time-coherence period as the transmission frame. The transmission frame is divided into two phases as the training phase and the data transmission phase. We consider  $T$ consecutive OFDM symbols which are equal to $M+1$ (number of sub-surfaces in the RIS +1) for the training phase since the RIS is fully passive and it does not have any active elements with transmit/receive RF chains.

\begin{figure}[ht]
	\centering
	\includegraphics[width=0.45\textwidth]{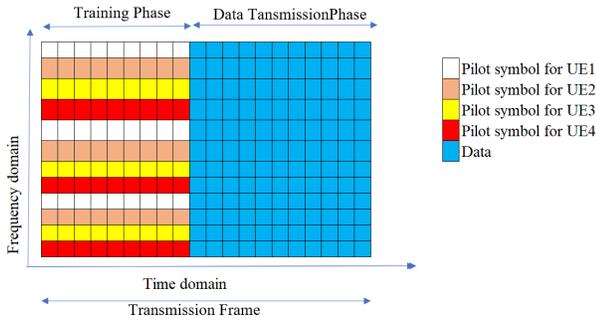}
	\caption{Illustration of the pilot allocation.}
	\label{fig:Illustration of the pilot allocation}
\end{figure}

Channel estimation is done by applying  pilot scheme as shown in Fig. \ref{fig:Illustration of the pilot allocation}.  In each OFDM symbol, there are $N_p$ number of pilot tones for each UE and the pilot set for $u$th UE is indexed as follows:
\begin{equation} \mathcal {P}_u=\left \{{0+(u-1), \Delta+(u-1),\ldots,(N_{p}-1)\Delta+(u-1) }\right \},\end{equation} 
where $\Delta =\lfloor N/N_{p}\rfloor$ denotes the frequency gap between adjacent pilots.  Therefore, $\Delta$ number of UEs' channel can be simultaneously estimated.

Let the received signal related to pilots $\mathcal {P}$ and $t$th OFDM symbol at  BS is ${\mathbf {Y}}_{\mathcal {P}}^{(t)}$. Hence the least-square estimation of the effective channel in $t$th OFDM symbol related to $u$th UE can be written as
\begin{align} \label{LSeq} {\mathbf {\hat{R}}_{u}^{(t,LS)}}=\left({\mathbf {X}_u}^{(t)} \right )^{-1}{\mathbf {Y}}_{\mathcal {P}}^{(t)} = \mathbf{H}_{\mathcal{P}_u} + \left({\mathbf {X}_u}^{(t)} \right )^{-1}\mathbf{V}_{\mathcal{P}}.\end{align}

We interpolate ${\mathbf {\hat{R}}_{u}^{(t,LS)}}$ along sub-carriers and obtain ${\mathbf {\Tilde {R}}_{u}^{(t)}}$  which can be written as follows:
\begin{equation} \label{rafterint} {{\mathbf {\Tilde R}}}_{u}^{(t)}={\mathbf {G}_{u}} { \boldsymbol \phi }^{(t)}+ {\mathbf {D}_{u}}+{\mathbf {\bar V}}^{(t)}={\vec {\mathbf {G}}_{u}} {\vec { \boldsymbol \phi }}^{(t)}+{\mathbf {\bar V}}^{(t)},\end{equation} 
where ${ \boldsymbol \phi }^{(t)}$ is the reflection coefficient vector at $t$th OFDM symbol,  ${\vec {\mathbf {G}_{u}}}=[{\mathbf {D}_{u}}, {\mathbf {G}_{u}}]$ and ${\vec { \boldsymbol \phi }}^{(t)}= \begin{bmatrix}1 & { \boldsymbol \phi }^{(t)}\end{bmatrix}^T$. 

By stacking all the OFDM symbols in training phase, $t=1,2,\ldots,M+1$, we can write the estimated channel matrix related to $u$th UE as, ${\mathbf {\hat H}_{u}}=[{\mathbf {\Tilde {R}}}_{u}^{(1)},{\mathbf {\Tilde R}}_{u}^{(2) },\ldots,{\mathbf {\Tilde R}}_{u}^{(M+1)}]$ and it is equivalent to
\begin{equation} \label{eq9} {\mathbf {\hat H}_{u}}={\vec {\mathbf {G}_{u}}} { {\boldsymbol{\vartheta }}}+{\mathbf {\bar V}},\end{equation}
where ${\boldsymbol{\vartheta }}=[{{\vec { \boldsymbol \phi }}^{(1)},{\vec { \boldsymbol \phi }}^{(2) },\ldots,{\vec { \boldsymbol \phi }}^{(M+1)}}]$ denotes the RIS reflection pattern matrix by collecting all reflection states during the training phase. We use a DFT-based reflection coefficient matrix during the training phase \cite{9053695}. Therefore, estimation of the direct channel and the cascaded channel due to the reflection of RIS,  related to $u$th UE becomes
\begin{equation}\label{eq10} \left [{{\mathbf {\hat D}_{u}}~{\mathbf {\hat G}_{u}} }\right]={\mathbf {\hat H}_{u}}{ {\boldsymbol{\vartheta }}}^{-1}.\end{equation}

\subsection{Channel Estimation -LMMSE}
\label{LMMSE}
MMSE is the optimal method of channel estimation. However, in a RIS-assisted system, it is difficult to obtain MMSE due to the non-linearity in the effective channel which is a combination of the direct UE-BS channel and the UE-RIS-BS cascade channels. Therefore, we present linear minimum mean square error (LMMSE) which is the best linear estimator that minimizes the mean square error (MSE). 

The LMMSE estimation of the effective channel  $\mathbf{H}_u$ in $n$th sub-carrier can be written as follows:
\begin{equation} \label{LMMSEeq}
    \mathbf{\hat {R}}_u^{LMMSE}(n) =  \mathbf{C}_u(n)\left(\mathbf{C}_u(n)+ \frac{\mathbf{I}_K}{SNR}\right)^{-1}\mathbf {\hat{R}}_{u}^{(LS)}(n),
\end{equation}
where $\mathbf{C}_u(n) = \mathbb{E}\left[\mathbf{H}_u(n)\mathbf{H}_u(n)^H\right]$ is the spatial correlation matrix of the effective channel of $u$th UE in $n$th sub-carrier and $SNR$ is the signal-to-noise ratio (SNR). Now $\mathbf{\hat {R}}_u^{LMMSE} $ can be interpolated along sub-carriers instead of ${\mathbf {\hat{R}}_{u}^{(LS)}}$ and follow the  steps mentioned in \eqref{rafterint}, \eqref{eq9}, \eqref{eq10} in the section \ref{LS} to find $[{{\mathbf {\hat D}_{u}}~{\mathbf {\hat G}_{u}} }]$. However, LMMSE has high complexity than the LS estimation since it has matrix inversions. 

\section{Untrained DNN for Channel Estimation}
\label{DNN}
The purpose of this paper is to investigate low complexity channel estimators to achieve high performance in terms of accuracy. We use an untrained DNN  to denoise the noisy LS estimation of the effective channel after interpolation ${\mathbf {\Tilde R}_{u}^{(t)}}$,  in order to enhance the performance of the channel estimation \cite{8949757}. The denoised effective channel is used to find the direct channel and the cascade channels with the help of the knowledge of DFT-based reflection pattern matrix. We are using an untrained DNN for denoising therefore it does not require training. Hence, complexity does not increase due to the addition of the denoising DNN.

The authors in \cite{lempitsky2018deep}, proposed a DNN design for image processing which does not require training, which is known as DIP. The authors in \cite{heckel2018deep}  further optimized the above algorithm and reduced the required number of parameters. The proposed DNN which is used to denoise the effective channel is based on the DIP model in \cite{heckel2018deep}.

The effective channel of $u$th UE after interpolation in \eqref{rafterint} is equivalent as follows when it is written in 3-dimensional form,
\begin{equation} \label{rfordl} \mathbf {R}_{DL} = \{\{\{{\Tilde{R}_{n,k,u}^{(t)}}\}_{k=1}^{K}\}_{n=1}^{N}\}_{t=1}^{T},\end{equation}
where $\Tilde{R}_{n,k,u}^{(t)}$ is the LS estimation of the effective channel after interpolation in the $t$th OFDM symbol, for $n$th subcarrier in the $k$th antenna related to $u$th UE. 
Tensors do not support complex operations, therefore the real and imaginary part of \eqref{rfordl} is separated into 2 independent channels, which mean the dimension of the input of the DNN becomes $\mathbf {R}_{DL} \in \mathbb{R}^{2K \times N \times T} $.

\begin{figure}[ht]
	\centering
	\includegraphics[width=0.5\textwidth]{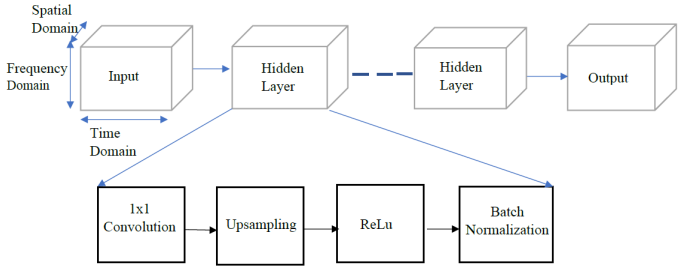}
	\caption{Illustration of the DNN.}
	\vspace{-2mm}
	\label{fig:Illustration of the DNN}
\end{figure}

The operating method of the DNN model is as follows:  First, we generate a random input tensor as the input and passed it through hidden layers. Initially, the weights of the DNN are randomly initialized. Then we optimize the weights of the hidden layers using  and updated them continuously. The hidden layers of the DNN model consist of four parts. They are  $1 \times 1$ convolution, an upsampler, a rectified linear unit (ReLU) activation function, and a batch normalization. The structure of the hidden layer is illustrated in Fig. \ref{fig:Illustration of the DNN}. However, the last hidden layer does not consist of an upsampler. 
\vspace{-3mm}

If the initial tensor is $\mathbf{R_0}$, then the tensor at the output layer becomes  $\mathbf {\hat {R}_{DL}} = f_{\theta_I}(f_{\theta_{I-1}}(...f_{\theta_0}(\mathbf{R}_0)))$. There are $I$ number of hidden layers and they are counted from $0$ to $I-1$, For the hidden layers, except for the last hidden layer,  $f_{\theta_i}$ can be written as
\vspace{-3mm}
\begin{multline}
 f_{\theta_i} = {BatchNorm(ReLU(Upsampler}(\theta_i \circledast \mathbf{R_i}))),\\ i=0,1, \cdots, I-2,\end{multline}
where $\theta_i$ is the parameters of the DNN and $\mathbf{r_i}$ is the tensor of the $i$th hidden layer. The last hidden layer is  $f_{\theta_{I-1}} = {BatchNorm(ReLU}(\theta_{I-1} \circledast \mathbf{R_{I-1}}))$ and the output layer is $f_{\theta_{I}} = \theta_{I} \circledast \mathbf{R_{I}}$. The parameters of the DNN is optimized as 
\begin{equation} \Theta ^{*} = \arg \min _\Theta ||\mathbf {R_{DL}}-\mathbf {\hat {R}_{DL}}||_{2}^{2},\end{equation} where $\Theta = (\theta _{0}, \theta _{1}, \cdots, \theta _{I})$. Final output of the DNN can be written as
\begin{equation} \mathbf { {R}_{DL}^{*}} = f_{\Theta ^{*}}(\mathbf{R}_0).\end{equation} 

Then, we follow  the  steps mentioned in \eqref{rafterint}, \eqref{eq9}, \eqref{eq10} in the section \ref{LS} with the denoised effective channel, $\mathbf { {R}_{DL}^{*}}$,  to find  $ [{{\mathbf {\hat D}_{u}}~{\mathbf {\hat G}_{u}} }]$.

 \subsection{Realistic System Model with Hardware Impairments}
 
 We consider two types of hardware impairments related to RIS-assisted systems.
 \begin{enumerate}
 \item The aggregate hardware impairments of transceiver: This can be modeled as independent additive distortion noise at the UE and the BS. During uplink channel estimation process, the distortion noise at $u$th UE is $\mathbf{\eta}_{u} \in \mathbb{C}^{N} \sim {\mathcal N_{c} }(\mathbf{0},{{\upsilon}}_u\mathbf{I}_{N})$ where   ${{\upsilon}}_u = \kappa_{UE}P_{u}$ and $P_u$ is the transmit power of the $u$th user. The distortion noise at the BS is $\mathbf{\eta}_{BS} \in \mathbb{C}^{N \times K} \sim {\mathcal N_{c} }(\mathbf{0},\mathbf{\Upsilon})$ where $\mathbf{\Upsilon} = \kappa_{BS}\sum _{u=0}^{ U}P_u\mathbf{I}_K \circ \mathbf{R}_u\mathbf{R}_u^H$. Here $\kappa_{UE}$ and $\kappa_{BS}$ is the  proportionality coefficients which characterize the levels of impairments at the UE and the BS respectively and $\circ$ is the Hadamard product. We assume $\kappa_{UE}$ is same for all UEs.
 
 \item The  hardware impairments of the RIS: This can be modeled as a phase noise. The phase noise of $m$th sub-surface of the RIS is $\Delta\varphi_m$ which is uniformly distributed in $[-\pi ,\pi)$. Then, $\Tilde{\phi}_m =  e^{-j(\varphi _{m}+\Delta\varphi_m})$.
 \end{enumerate}
 
 The received signal at the BS with hardware impairments is as follows:
 \begin{equation} \label{eqhwi}{\mathbf{\Tilde{Y}}}=\sum _{u=1}^{U} \left({\mathbf{X_u}+ \text{diag}(\mathbf{\eta}_u) }\right) \left({ \mathbf{G}_u \mathbf{\Tilde{\phi}}  + {\mathbf{D}}_{u}}\right)+ \mathbf{\mathbf{\eta}_{BS}}+{\mathbf{V}},\end{equation}
 where $\mathbf{\Tilde{\phi}} = [\Tilde{\phi _{1}},\Tilde{\phi _{2}}, \ldots, \Tilde{\phi _{M}}]$. The LS estimation of the effective channel in $t$th OFDM symbol related to $u$th UE when hardware impairments are present can be written as

\begin{footnotesize}
\begin{align} {\mathbf {\hat{S}}_{u}^{(t,LS)}}=&\left({\mathbf {X}_u}^{(t)} \right )^{-1}{{\mathbf {\Tilde{Y}}}}_{\mathcal {P}}^{(t)} \notag  = \mathbf{H}_{\mathcal{P}_u} +\left({\mathbf {X}_u}^{(t)} \right )^{-1}\sum _{i=1}^{U}{ \text{diag} (\mathbf{\eta}_i)} \mathbf{H}_{\mathcal{P}_i}\notag \\+& \left({\mathbf {X}_u}^{(t)} \right )^{-1}\mathbf{\mathbf{\eta}_{BS}} + \left({\mathbf {X}_u}^{(t)} \right )^{-1}\mathbf{V}_{\mathcal{P}}.\end{align}
\end{footnotesize}

The conventional distortion aware LMMSE estimation in $n$th sub-carrier when there is hardware impairments can be written as follows:

\vspace{-2mm}
\begin{align}
    \mathbf{\hat {S}}_u^{LMMSE}(n) =  \mathbf{C}_u(n)\Lambda(n)^{-1}\mathbf {\hat{S}}_{u}^{(LS)}(n), 
\end{align}
where $ \Lambda(n) = \mathbf{C}_u(n)+ \kappa_{UE}\sum _{j=1}^{U}\mathbf{C}_j(n) + \kappa_{BS}\sum _{j=1}^{U} \mathbf{I}_K \circ \mathbf{C}_j(n)  \frac{\mathbf{I}_K}{SNR}$. Then   $ \mathbf{\hat {S}}_u^{LMMSE}$ can be interpolated along sub-carriers and follow the  steps mentioned in \eqref{rafterint}, \eqref{eq9}, \eqref{eq10} in the section  \ref{LS}   to find $ [{{\mathbf {\hat D}_{u}}~{\mathbf {\hat G}_{u}} }]$.

However, finding the distortion aware LMMSE estimations in practice  is very challenging, since for that the knowledge of distortion parameters is necessary.  Therefore, we can denoise  the interpolated $\mathbf {\hat{S}}_{u}^{(LS)}$ using our proposed DNN mentioned in the beginning of this section and obtain more accurate estimation for the direct channel and cascade channel without any knowledge of distortion parameters.

\section{Numerical Results }
\label{results}
 In this section, we present our simulation results to show the performance of our proposed technique. In our simulations, we set the number of  UEs, $U=4$ and  number of antennas in BS  $K=32$.  RIS is consists with reflecting elements  with half-wavelength spacing and we set the number of elements in the RIS as, $M \times M = \Tilde{M}=15 \times 15 = 225$. We divide the RIS into 15 sub-surfaces by grouping adjacent elements which have high correlation.  We assume that one transmission frame consists of 140 OFDM symbols and out of that, 16 OFDM symbols are taking for the training phase. There are 64  sub-carriers in each OFDM symbol. The DNN model has 6 hidden layers, i.e., $I =6$. We use Adam optimization with a learning rate of 0.01 to optimize the parameters in the DNN. 

\begin{figure}[ht]
	\centering
	\includegraphics[width=0.4\textwidth]{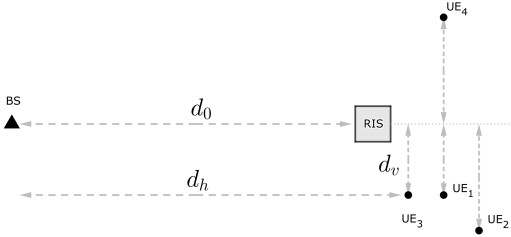}
	\caption{Simulation setup.}
	\vspace{-2mm}
	\label{fig:uedistance}
\end{figure}

We assume that the UEs lie near to the RIS as shown in Fig. \ref{fig:uedistance}. The horizontal distance between the BS and the RIS is $d_0=50m$. 
\vspace{-4mm}

Horizontal and vertical distance parameters of UEs' locations are mentioned in  Table \ref{table:distance parameters}. The UE-BS distances and the UE-RIS distances are calculated using $d_{UE-BS} = \sqrt{d_h^2 + d_v^2}$ and $d_{UE-RIS} = \sqrt{(d_h-d_0)^2+d_v^2}$, respectively.

	\vspace{-3mm}
\begin{table}[h]
	\caption{Distance Parameters} 
	\centering 
	\begin{tabular}{|c| c|c|c|c| } 
		\hline 
	    	& UE1 & UE2& UE3&UE4\\
		\hline
		$d_v (m)$& 2 &3&2&3\\
	    \hline
		$d_h (m)$&52&53&51&52\\
		\hline
	\end{tabular}
	\label{table:distance parameters} 
		\vspace{-2mm}
\end{table}

 The Zadoff-Chu sequence is used as the pilot sequence during the training phase.  Time-domain channels in all the links are modeled as spatially correlated Raleigh channels. The spatial correlation of the RIS is calculated according to \cite{bjornson2020rayleigh}. Further, we consider the Toeplitz-structured correlation with a 0.7 correlation factor at the BS. The path loss is calculated based on the 3GPP Urban Micro (UMi) scenario from \cite{3gpp} for a carrier frequency of 6 GHz. Number of delay taps in each link are  $L_d=6$, $L_q=2$ and $L_b=5$. We assume that transmit power at each UE is 20dBm.  

\begin{figure}[ht]
	\centering
	\includegraphics[width=0.5\textwidth]{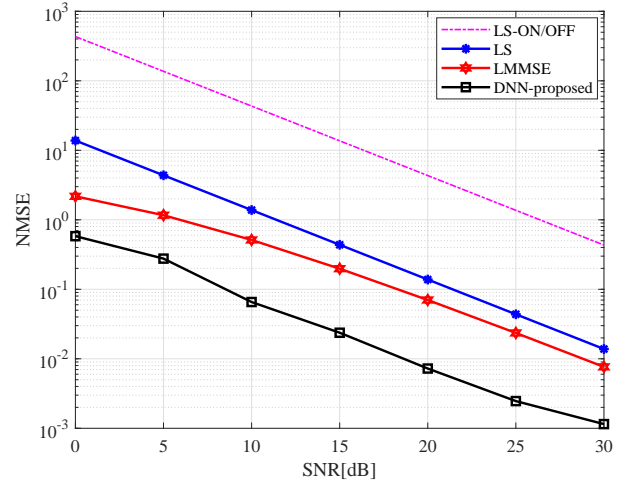}
	\caption{Comparison the proposed method with existing methods  when Np = 16.}
	\vspace{-3mm}
	\label{fig:compare methods}
\end{figure}

We use normalized mean square (NMSE) as the performance metric which can be calculated as 
$ \text {NMSE} = \mathbb {E}\left [{\frac {||\mathbf { {H}_{actual}}-\mathbf { {\hat {H}}_{estimated}}||_{2}^{2}}{||\mathbf { {H}_{actual}}||_{2}^{2}}}\right]. $
The Fig. \ref{fig:compare methods} shows the comparison of  NMSE of the proposed estimator with other methods  against SNR. For comparison we adopt the ON/OFF LS channel estimation for RIS-assisted OFDM system which is proposed by the authors in \cite{9039554}. Further we compare our proposed method based on the DNN mentioned in section \ref{DNN} with the conventional channel estimation methods mentioned in sections \ref{LS}, \ref{LMMSE}. In this scenario we obtain result for ideal system without hardware impairments. The Fig. \ref{fig:compare methods} concludes that our proposed method based on the DNN achieves highest performance in terms of accuracy than the other methods.

 By comparing \eqref{LSeq} and \eqref{LMMSEeq} we can say that LS estimation has low complexity than the LMMSE since LMMSE estimation requires the knowledge of the correlation matrix of the channel and also it has matrix inversion. According to Fig. \ref{fig:compare methods} LMMSE estimation is more accurate than LS estimation but that performance increment comes at a cost of increased complexity. However, our proposed technique mentioned in section \ref{DNN}, has a complexity equal to LS estimation since it uses only LS estimations and DNN  does not incur any complexity increase due to its untrained nature. It implies that our proposed method can achieve high accuracy with low complexity compared to existing methods. 

\begin{figure}[ht]
	\centering
	\includegraphics[width=0.5\textwidth]{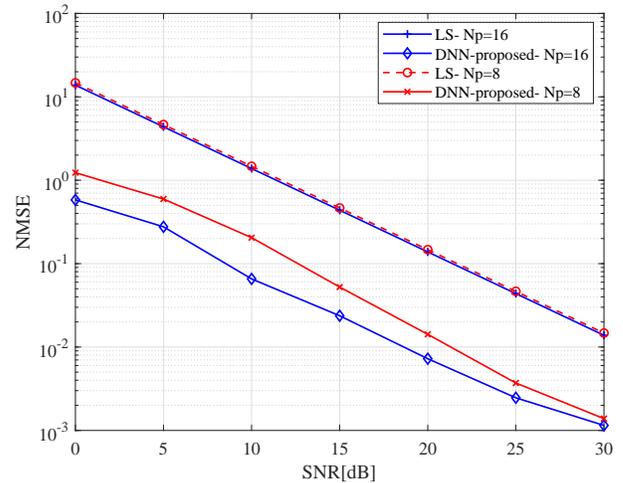}
	\caption{NMSE of the proposed estimator with proposed DNN  for different $N_p$ with respect to SNR.}
	\vspace{-2mm}
	\label{fig:Npgraph}
\end{figure} 

In the simulations, we use 4 UEs and 64  sub-carriers for one OFDM symbol. Therefore the maximum pilot length for one UE is 16. Fig. \ref{fig:Npgraph} shows the results for the simulations  for both the scenarios $N_p=8 $ and $N_p=16$ for the both methods mentioned in sections \ref{LS} and \ref{DNN}. We obtain the results in Fig. \ref{fig:Npgraph} for ideal system without hardware impairments. It shows that when the pilot length increases the accuracy also increases.

 \begin{figure}[ht]
	\centering
	\includegraphics[width=0.5\textwidth]{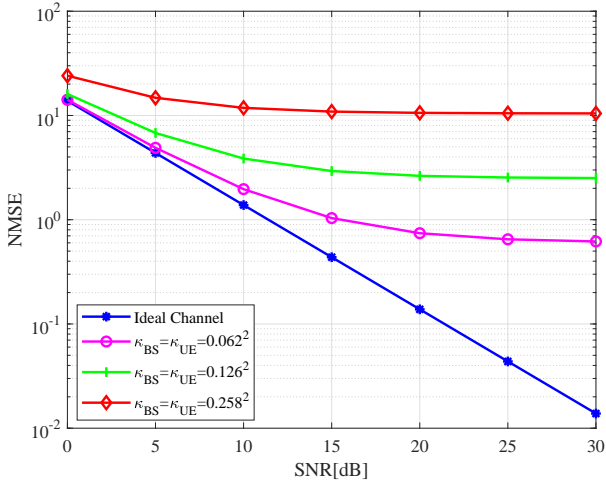}
	\caption{NMSE against SNR  for different $\kappa_{UE}$ , $\kappa_{BS}$ combinations and the ideal channel. }
	\vspace{-2mm}
	\label{fig:hwicomparison}
\end{figure} 

 \begin{figure}[ht]
	\centering
	\includegraphics[width=0.5\textwidth]{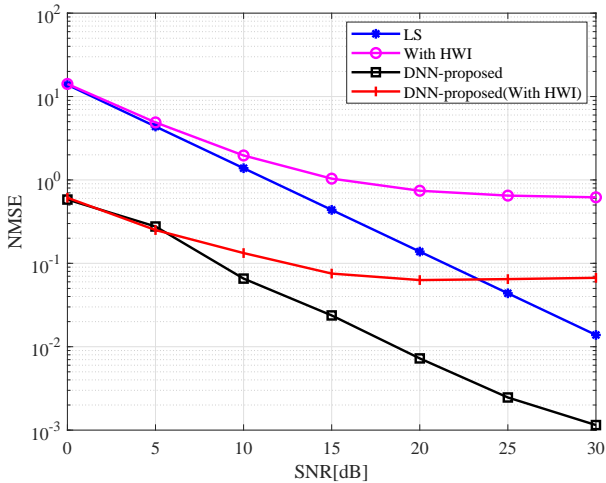}
	\caption{Comparison of with and without hardware impairments (HWI) with the proposed DNN when $\kappa_{BS}=\kappa_{UE}=0.05^2$ }
	\vspace{-2mm}
	\label{fig:hwi}
\end{figure} 

Distortion at the BS and UE can be written as $\kappa_{BS}= \kappa_{UE} = 2^{-2b}/(1-2^{-2b})$ \cite{papazafeiropoulos2021intelligent}. We set $\kappa_{BS} =\kappa_{UE}= 0.258^2, 0.126^2$ and $0.062^2$ for $b = 2,3$ and $4$ bits respectively. We present the variation of the NMSE for different  $\kappa_{UE}$ and $\kappa_{BS}$ values  with respect to the SNR in Fig. \ref{fig:hwicomparison}, when $N_p=16$.  It shows when the $\kappa_{UE}$ and $\kappa_{BS}$ values increase, which means the distortion at the transceiver increases,   the accuracy of the estimation decreases. We simulate the system with hardware impairments with the proposed DNN method. Fig. \ref{fig:hwi} shows the comparison of the accuracy of the RIS-assisted system with hardware impairments when we use the conventional methods and our proposed method. The results show our proposed method can achieve high accuracy even when there is interference caused by the hardware impairments. This concludes that our method is robust to systems with hardware impairments.

\section{Conclusion}
\label{conclusion}
In this paper, we have proposed a low complexity channel estimator of RIS-assisted multi-user SIMO OFDM system based on a denoising DNN. Our proposed  estimator is robust to the hardware impairments at the transceiver and the RIS. We have used an untrained DNN to denoise the effective channel obtained from the conventional pilot-based LS  estimation. This denoised effective channel is used to find the estimations for direct channel and cascade channels. Furthermore, the proposed deep-learning based estimator is flexible for the changes in the environment and the channel since it is doing calculations on the fly only with the help of the pilot-symbols and the received signal. Numerically, we have shown that our proposed method has high performance in terms of accuracy than the conventional methods even when the system has hardware impairments. 


\bibliographystyle{IEEEbib}

\bibliography{di}

\begin{thebibliography}{10}

\bibitem{9326394}
Q.~{Wu}, S.~{Zhang}, B.~{Zheng}, C.~{You}, and R.~{Zhang},
\newblock ``{Intelligent Reflecting Surface Aided Wireless Communications: A
  Tutorial},''
\newblock {\em IEEE Transactions on Communications}, pp. 1--1, 2021.

\bibitem{9039554}
Y.~{Yang}, B.~{Zheng}, S.~{Zhang}, and R.~{Zhang},
\newblock ``{Intelligent Reflecting Surface Meets OFDM: Protocol Design and
  Rate Maximization},''
\newblock {\em IEEE Transactions on Communications}, vol. 68, no. 7, pp.
  4522--4535, 2020.

\bibitem{8683663}
D.~{Mishra} and H.~{Johansson},
\newblock ``{Channel Estimation and Low-complexity Beamforming Design for
  Passive Intelligent Surface Assisted MISO Wireless Energy Transfer},''
\newblock in {\em ICASSP 2019 - 2019 IEEE International Conference on
  Acoustics, Speech and Signal Processing (ICASSP)}, 2019, pp. 4659--4663.

\bibitem{9090876}
A.~M. {Elbir}, A.~{Papazafeiropoulos}, P.~{Kourtessis}, and S.~{Chatzinotas},
\newblock ``{Deep Channel Learning for Large Intelligent Surfaces Aided mm-Wave
  Massive MIMO Systems},''
\newblock {\em IEEE Wireless Communications Letters}, vol. 9, no. 9, pp.
  1447--1451, 2020.

\bibitem{9053695}
T.~L. {Jensen} and E.~{De Carvalho},
\newblock ``{An Optimal Channel Estimation Scheme for Intelligent Reflecting
  Surfaces Based on a Minimum Variance Unbiased Estimator},''
\newblock in {\em ICASSP 2020 - 2020 IEEE International Conference on
  Acoustics, Speech and Signal Processing (ICASSP)}, 2020, pp. 5000--5004.

\bibitem{8937491}
B.~{Zheng} and R.~{Zhang},
\newblock ``{Intelligent Reflecting Surface-Enhanced OFDM: Channel Estimation
  and Reflection Optimization},''
\newblock {\em IEEE Wireless Communications Letters}, vol. 9, no. 4, pp.
  518--522, 2020.

\bibitem{papazafeiropoulos2021intelligent}
Anastasios Papazafeiropoulos, Cunhua Pan, Pandelis Kourtessis, Symeon
  Chatzinotas, and John~M Senior,
\newblock ``{Intelligent Reflecting Surface-assisted MU-MISO Systems with
  Imperfect Hardware: Channel Estimation, Beamforming Design},''
\newblock {\em arXiv preprint arXiv:2102.05333}, 2021.

\bibitem{8949757}
E.~{Balevi}, A.~{Doshi}, and J.~G. {Andrews},
\newblock ``{Massive MIMO Channel Estimation With an Untrained Deep Neural
  Network},''
\newblock {\em IEEE Transactions on Wireless Communications}, vol. 19, no. 3,
  pp. 2079--2090, 2020.

\bibitem{lempitsky2018deep}
Victor Lempitsky, Andrea Vedaldi, and Dmitry Ulyanov,
\newblock ``{Deep image prior},''
\newblock in {\em 2018 IEEE/CVF Conference on Computer Vision and Pattern
  Recognition}. IEEE, 2018, pp. 9446--9454.

\bibitem{heckel2018deep}
Reinhard Heckel and Paul Hand,
\newblock ``{Deep decoder: Concise image representations from untrained
  non-convolutional networks},''
\newblock {\em arXiv preprint arXiv:1810.03982}, 2018.

\bibitem{9366894}
N.~K. {Kundu} and M.~R. {McKay},
\newblock ``{Channel Estimation for Reconfigurable Intelligent Surface Aided
  MISO Communications: From LMMSE to Deep Learning Solutions},''
\newblock {\em IEEE Open Journal of the Communications Society}, vol. 2, pp.
  471--487, 2021.

\bibitem{bjornson2020rayleigh}
Emil Bj{\"o}rnson and Luca Sanguinetti,
\newblock ``{Rayleigh fading modeling and channel hardening for reconfigurable
  intelligent surfaces},''
\newblock {\em IEEE Wireless Communications Letters}, 2020.

\bibitem{3gpp}
3GPP,
\newblock ``{TR 38.901 V16.1.0 - Study on channel model for frequencies from
  0.5 to 100 GHz},''
\newblock Tech. {R}ep., 3GPP, 2020.

\end{thebibliography}

\end{document}